\documentstyle[12pt, epsf]{article}
\newcommand{\bt}{{\tilde{\beta}}}

\newcommand{\ks}{{K^{\ast}}}

\newcommand{\kt}{{K_2^{\ast}(1430)}}
\newcommand{\kol}{{K_1(1270)}}
\newcommand{\koh}{{K_1(1400)}}
\newcommand{\kz}{{K_0^{\ast}(1430)}}

\newcommand{\ksl}{{K^{\ast}(1410)}}
\newcommand{\ksh}{{K^{\ast}(1680)}}
\newcommand{\k}{{K(1460)}}

\newcommand{\ktl}{{K_2(1770)}}
\newcommand{\kth}{{K_2(1820)}}
\newcommand{\kf}{{K_3^{\ast}(1780)}}

\newcommand{\dex}{{D(2637)}}
\newcommand{\ds}{{D^{\ast}}}
\newcommand{\dpp}{{D^{'}}}
\newcommand{\dsp}{{{D^{\ast}}^{'}}}

\newcommand{\dts}{{D_2^{\ast}(2460)}}
\newcommand{\doo}{{D_1(2420)}}
\newcommand{\dox}{{D_{1\; j_q=\frac{1}{2}}^{\ast}}}
\newcommand{\dz}{{D_0^{\ast}}}

\newcommand{\df}{{D_3^{\ast}}}
\newcommand{\dt}{{D_{2\; j_q=\frac{5}{2}}}}
\newcommand{\dtx}{{D_{2\; j_q=\frac{3}{2}}}}
\newcommand{\dos}{{D_{1 \; j_q=\frac{3}{2}}}}

\newcommand{\des}{{D_s}}
\newcommand{\dss}{{D_s^{\ast}}}

\newcommand{\bex}{{B(5905)}}
\newcommand{\bs}{{B^{\ast}}}
\newcommand{\btab}{\begin{tabbing}}
\newcommand{\etab}{\end{tabbing}}

\newcommand{\beqn}{\begin{equation}}
\newcommand{\eeqn}{\end{equation}}
\newcommand{\barr}[1]{\begin{array}{#1}}
\newcommand{\earr}{\end{array}}
\newcommand{\beqna}{\begin{eqnarray}}
\newcommand{\eeqna}{\end{eqnarray}}
\newcommand{\btablec}{\begin{table} \begin{center}}
\newcommand{\etablec}{\end{center} \end{table}}

\newcommand{\gapproxeq}{\lower.7ex\hbox{$\;\stackrel{\textstyle>}
{\sim}\;$}}
\newcommand{\plabel}[1]{\label{#1}}
\newcommand{\pbibitem}[1]{\bibitem{#1}}
\input epsf

\begin{document}
\title{
\begin{flushright} 
\small{hep-ph/9809575} \\ 
\small{LA-UR-98-4299} 
\end{flushright} 
\vspace{0.6cm}  
\Large\bf Interpretation of $\dex$ from heavy quark symmetry}
\vskip 0.2 in
\author{Philip R. Page\thanks{\small \em E-mail:
prp@lanl.gov} \\
{\small \em  Theoretical Division, Los Alamos
National Laboratory,}\\ 
{\small \em P.O. Box 1663, Los Alamos, NM 87545, USA}}
\date{September 1998}
\date{}
\maketitle
\begin{abstract}{We demonstate from heavy quark symmetry that the width of
$\dex$ claimed by the DELPHI Collaboration is inconsistent with any bound 
state with one charm quark predicted in the $\dex$ mass region, except 
possibly $\df$, $\dt$ or $\dpp$. The former two possibilities are favoured by
heavy quark mass relations.}
\end{abstract}
\bigskip


PACS number(s): 11.30.Hv \hspace{.2cm}13.25.Es  \hspace{.2cm}13.25.Ft 
\hspace{.2cm}14.40.Ev\hspace{.2cm}14.40.Lb

\section{Introduction}

The DELPHI Collaboration recently presented evidence for a new state $\dex$
at $2637\pm 2\pm 6$ MeV with a width of $< 15$ MeV at 95\% confidence 
\cite{delphi}.
A signal of $66 \pm 14$ events was detected, a $4.7\sigma$ effect.
The state was observed decaying to $\ds^{+}\pi^+\pi^-$. 
However, its existence has not been confirmed by the CLEO and
OPAL Collaborations in the same decay channel \cite{opal}. Moreover,
there is no evidence for $\dex$ in $\ds\pi$ \cite{delphi,opal}. 

In this Report we shall assume the validity of the DELPHI claim, and provide
theoretical interpretations for $\dex$, based on two assumptions:

\begin{enumerate}

\item The validity of lowest order heavy quark symmetry decay relations between
the $K$--meson and corresponding $D$--meson systems.

\item The validity of experimental data on the established $K$--mesons 
$\ksl,\;\ksh,$ $\kz,\; \koh$ and $\kf$ \cite{disc}.
 
\end{enumerate}

These are the {\it only} assumptions that will be made, unless otherwise 
indicated.
The latter assumption is inevitable
given that no new data is likely to be forthcoming soon.
The former assumption is needed to make extrapolations, independent of 
detailed dynamical models, from known
$K$--meson decays to unknown $D$--meson decays. Although not always good 
within the 
simplest form factor assumptions (see below), it has been argued to be 
qualitatively valid \cite{suzuki} in an analysis of 
$\kol$ and $\kt$ \cite{eichten93,eichten94}. A specific example of how 
heavy quark symmetry for strange quarks gives predictions in the right 
ballpark is as follows. The D-- to S--wave width ratio for 
$\koh\rightarrow\ks\pi$ is $0.04\pm 0.01$ \cite{pdg98} 
(heavy quark symmetry predicts zero \cite{suzuki}); and for 
$\kol\rightarrow\ks\pi$ is $1.0\pm 0.7$  \cite{pdg98} 
(heavy quark symmetry predicts infinity \cite{suzuki}).

The radial excitation of the $\ds$, referred to as $\dsp$, is predicted  
to have a mass of
2640 MeV in 
a model which predicted the $\dts$ mass within 40 MeV of experiment 
\cite{isgur85};
and 2629 MeV in a recent model which is in agreement within
20 MeV for the observed charm orbital states \cite{ebert}.
The noticeable proximity of these potential model predictions to the 
mass of $\dex$ leads 
DELPHI to identify it as $\dsp$, and hence as $J^P=1^-$. However, 
there are several experimentally unobserved conventional mesons which can also 
decay to $\ds \pi\pi$
and are expected in potential models in the vicinity of $\dex$. Of these, the
radially excited $D$ with $J^P=0^-$, referred to as $\dpp$, 
would be nearest in mass at 
2580 MeV \cite{isgur85} or 2579 MeV \cite{ebert}. 
Next nearest would be the $\dox$, a possibly observed \cite{cleo} $1^+$ at
 2460 MeV \cite{isgur85}, 2501 MeV \cite{ebert}  or 2585 MeV \cite{isgur98}; 
and
the $0^+$ $\dz$ at 2400 MeV \cite{isgur85}, 2438 MeV \cite{ebert} or 
2554 MeV \cite{isgur98}.
The $1^-$ $\dos$ is at 2820 MeV \cite{isgur85}.
The $3^-$ $\df$ should be at
2830 MeV \cite{isgur85} or $2760\pm 70$ MeV \cite{nogteva}. The $2^-$ states
$\dt$ and $\dtx$ are within 20 MeV of $\df$ \cite{isgur85}. 
Given the small error bars of the DELPHI mass measurement, all 
interpretations except $\dsp$ fail on mass grounds. 

The purpose of this Report is to check which of the preceding possibilities 
can reproduce the tiny total width of $<15$ MeV claimed by DELPHI, 
assuming them 
to have the mass of $\dex$. The masses of all experimentally
known states will be taken from the PDG \cite{pdg98}.

For a given heavy--light meson with total angular momentum $\vec{J}$, 
let $\vec{s}_Q$ ($s_Q = \frac{1}{2}$) be the spin of the heavy quark
and $\vec{\j}_q$ the total angular momentum  of the light degrees of 
freedom.
Consider the decay of a heavy--light meson characterized by 
$J,\; j_q$ to an outgoing heavy--light meson characterized by 
$J^{'},\; j_q^{'}$ and a light meson with spin ${s}_h$. 
The light meson has  
orbital angular momentum $\ell$ relative to the outgoing heavy--light meson.
The decay amplitude  
satisfies certain symmetry relations because the decay dynamics become 
independent
of the heavy quark spin in the heavy quark limit of QCD \cite{wise}. 
The two--body decay
width can be factored into a reduced form factor multiplied by a 
normalized 6--$j$ symbol
 \cite{wise}

\beqn \plabel{hq}
\Gamma = \left(\sqrt{(2 J^{'}+1)(2 j_q +1)} 
\left\{ \barr{ccc}  s_Q    &  j_q^{'}    &   J^{'}     \\
                   j_h     &  J    &  j_q   \earr \right\}  \right)^2 
\; p^{2\ell+1}\; F^{j_q\; j_q^{'}}_{j_h\; \ell} (p^2)
\eeqn
where $\vec{\j}_h \equiv \vec{s}_h + \vec{\ell}$. The 6--$j$ symbols are 
evaluated in ref.
\cite{var}. $p$ is the magnitude of the three--momentum of the decay 
products in the rest
frame of the initial state. Eq. \ref{hq} neglects corrections to the heavy 
quark limit,
except in as far as they modify $p$.
One essential idea of the heavy quark limit is that the spin of the 
heavy quark and the total angular momentum of the light degrees of freedom 
are seperately conserved \cite{wise},
i.e. $\vec{\j}_q = \vec{\j_q^{'}} + \vec{\j}_h$. This conservation law is 
in addition
to the usual conditions of conservation of total angular momentum 
$\vec{J} = \vec{J^{'}}+\vec{\j}_h$
and parity. For the remainder of this Report we shall restrict to $\ell$ 
allowed by all these
conservation conditions. Heavy quark symmetry does not predict the magnitude 
and functional
dependence of the reduced form factor
$F^{j_q\; j_q^{'}}_{j_h\; \ell} (p^2)$ for a particular decay. Once 
determined from
experimentally well established decays of $K$--mesons with given 
$j_q,\; j_q^{'}$, 
this quantity may be used to predict related decays of
both $K$-- and $D$--mesons with the same $j_q,\; j_q^{'}$.
 
\section{Interpreting $\dex$: Gaussian form factor}

\begin{table}[t]
\begin{center}
\caption{\small Widths of $\dex$ to $D\pi$ and $D^{\ast}\pi$ in MeV. 
The interpretation and $j_q$ of $\dex$ is
given in the first and third columns respectively. Blank entries are 
identical to those
above them. Since there are reasons to doubt that $\ksl$ is the radially 
excited $\ks$, 
the $\ksh$ is often taken to be the radial excitation 
\protect\cite{pdg98,burakovsky97}.
Another quark model interpretation of $\ksh$ is as a D--wave meson 
\protect\cite{pdg98,burakovsky97}, so that
$j_q = \frac{3}{2}$. 
The only interpretation of $\kz$ is as a P--wave meson \protect\cite{pdg98},
so that $j_q=\frac{1}{2}$.  }
\begin{tabular}{|c|l|c|c|r|r|}
\hline 
$\dex$ & K--meson data used \cite{tot} & $j_q$ & Form Factor & $D\pi$ & 
$\ds\pi$  \\
\hline 
$\dsp$ & $\Gamma(\ksl \rightarrow K\pi) = 15$ MeV     & $\frac{1}{2}$ & 
$F^{\frac{1}{2}\frac{1}{2}}_{1\; 1}(0)$ & 17 & 22 \\
       & $\Gamma(\ksh \rightarrow K\pi) = 125$ MeV    &               & 
                                     & 86 & 115 \\
       & $\Gamma(\ksh \rightarrow \ks\pi) = 96.3$ MeV &               &  
                                    & 53 & 71\\
$\dpp$ & $\Gamma(\ksl \rightarrow K\pi)$              &               &  
                                    & - & 33 \\
       & $\Gamma(\ksh \rightarrow K\pi)$              &               &  
                                    & - & 172  \\
       & $\Gamma(\ksh \rightarrow \ks\pi)$            &               & 
                                     & - & 106 \\
$\dz$  & $\Gamma(\kz\rightarrow K\pi) = 270$ MeV      &               & 
$F^{\frac{1}{2}\frac{1}{2}}_{0\; 0}(0)$ & 270 & - \\
$\dox$ &                                               &               &  
                                    & - & 260 \\
$\dos$ & $\Gamma(\ksh \rightarrow K\pi)$               & $\frac{3}{2}$ & 
$F^{\frac{3}{2}\frac{1}{2}}_{1\; 1}(0)$ & 87 & 29 \\
       & $\Gamma(\ksh \rightarrow \ks\pi)$             &               & 
                                     & 213 & 71 \\
$\dtx$ & $\Gamma(\ksh \rightarrow K\pi)$               &               & 
                                     & - & 86 \\
       & $\Gamma(\ksh \rightarrow \ks\pi)$             &               &  
                                    & - & 212 \\
\hline 
\end{tabular}
\end{center}
\end{table}

We shall assume a Gaussian form for the reduced form factor \cite{eichten93}

\beqn \plabel{hqs}
 F^{j_q\; j_q^{'}}_{j_h\; \ell} (p^2) = F^{j_q\; j_q^{'}}_{j_h\; \ell} (0)\;
\exp(-\frac{p^2}{6 \beta^2})
\eeqn
in this section. The Gaussian form arises in decay models where simple
harmonic oscillator wave functions are used \cite{godfrey91,kokoski87,biceps},
and the value $\beta=0.4$ GeV is phenomenologically successful
\cite{eichten93,eichten94,godfrey91,kokoski87,biceps}. We shall adopt this
value, although our predictions are stable under the variation
 $\beta = 0.35 - 0.45$ GeV.
Tables 1 and 2 indicate the interpretations of $\dex$ that will be explored.

\begin{table}[p]
\begin{center}
\caption{\small Partial widths of $\df$ and $\dt$ in MeV. Blank entries are
identical to those above them. In some form factors we have explicitly 
indicated the light meson $\eta$ or $\rho$, in 
order to distinguish them from form factors for $\pi$.
The only quark model interpretation of $\kf$ is as a D--wave meson
\cite{pdg98,burakovsky97}, so that $j_q = \frac{5}{2}$. 
Decays of $\dex$ to $D\omega,\; D^{\ast}\rho$ are below threshold by
more than half a width of $\omega$ and $\rho$ respectively, and are not 
calculated in this
Report. However, $\dt\rightarrow D\omega$ and 
$\df,\;\dt\rightarrow D^{\ast}\rho$ can be 
in P--wave and hence competitive with the rates in the text, although current 
experimental data on K--mesons do not give sufficient information to 
estimate these rates from heavy quark
symmetry. $\dt\rightarrow\dz\pi$ is a D--wave decay and
$\df,\;\dt\rightarrow\dox\pi$ a D--wave decay at threshold, using the 
$\dox$ and $\dz$ masses of
ref. \cite{ebert}. These decays cannot be estimated from
experimental data.
$\dagger$ Assuming SU(3) symmetry. 
$\ddagger$ This is an F--wave decay
at threshold, and hence very sensitive to phase space. We smear the partial 
width
(Eqs. \protect\ref{hq} and \protect\ref{hqs}) over a relativistic 
Breit--Wigner form
to take account of the 150 MeV width of the $\rho$. $\clubsuit$ This 
decay involves
form factors which cannot be estimated  from experimental
data. $\amalg$ The width of $\dts$ has been smeared over. }
\begin{tabular}{|l|c|c|r|r|}
\hline 
$K$--meson data used \protect\cite{tot} & Form Factor & Decay Mode &$\df$ 
& $\dt$ \\
\hline 
$\Gamma(\kf \rightarrow K\pi) = 29.9$ MeV & 
$F^{\frac{5}{2}\frac{1}{2}}_{3\; 3}(0)$     & $D\pi$        & 7.8 & -  \\
                                          &                  
                        & $\ds\pi$      & 3.4 & 5.9\\
$\Gamma(\kf \rightarrow \ks\pi)= 32 $ MeV &                  
                        & $D\pi$        & 22  & -  \\
                                          &                     
                     & $\ds\pi$      & 7.8 & 17 \\
$\Gamma(\kf \rightarrow K\pi,\ks\pi)$     &                       
                   & $\des K$ $\dagger$     &$<.6$& -  \\
                                          &                        
                  & $\dss K$ $\dagger$     &$\sim 0$&$\sim 0$ \\
$\Gamma(\kf \rightarrow K\eta) = 48$ MeV  
& $F^{\frac{5}{2}\frac{1}{2}\;\eta}_{3\; 3}(0)$ & $D\eta$       & 2.9 & -  \\
                                          &                        
                  & $\ds\eta$     & 0.1 &  0.2 \\
$\Gamma(\kf \rightarrow K\rho) = 49$ MeV  
& $F^{\frac{5}{2}\frac{1}{2}\;\rho}_{3\; 3}(0)$ & $D\rho$  $\ddagger$ 
   & 0.7 &  $\clubsuit$ \\
$\Gamma(\kf\rightarrow\kt\pi) < 25$ MeV   
& $F^{\frac{5}{2}\frac{3}{2}}_{2\; 2}(0)$     & $\dts\pi$ $\amalg$     
&$<0.5$& $< 0.2$ \\
                                          &             
                             & $\doo\pi$     &$<0.2$&$<1.1$ \\
\hline 
\end{tabular}
\end{center}
\end{table}

The first entry in Table 1 will be discussed 
in detail to clarify the methods used.
For our heavy quark symmetry analysis it is not neccesary
to know the nature of $\ksl$, only the value of $j_q$, which can be
$\frac{1}{2}$ or $\frac{3}{2}$ since $J=1$. We shall motivate
our choice of $j_q$ from the known quark model interpretation.
The only interpretation of $\ksl$ is as a radially excited
$\ks$ \cite{pdg98,burakovsky97}, so that $j_q = \frac{1}{2}$. 
However, $\ksl$ may have a non--conventional--meson component, e.g. a 
low--lying
$1^-$ hybrid meson with $j_q=\frac{1}{2}$.
Noting that the $\pi$ has $s_h=0$,
we deduce from Eqs. \ref{hq} and \ref{hqs}, using the experimental data on
$\Gamma(\ksl \rightarrow K\pi)$,  the value of 
$F^{\frac{1}{2}\frac{1}{2}}_{1\; 1}(0)$. From this
 $\Gamma(\dsp\rightarrow D\pi,\; \ds\pi)$
is calculated. The total width of $\dsp$ is found to be appreciably higher 
than the DELPHI value.
The same holds true for all other possibilities explored in Table 1.

$\kf$  has been used in an
analogous study to the one in this 
Report \cite{eichten93,eichten94}.
There the heavy quark symmetry partners $\df$ and $\dt$ have been found to 
be 193 MeV and 99 MeV 
wide \cite{eichten94}, respectively,
due to the high mass of 2830 MeV used \cite{hill94}. In this work 
we use the mass
of the $\dex$ by fiat, so that the total widths should be 
substantially smaller. 

The partial widths of $\df$ are estimated in Table 2. 
All decay modes other than $D\pi$ and
$D^{\ast}\pi$ contributes $4-5$ MeV. The $D\pi$ and $\ds\pi$ partial widths
depend on which K--meson decay they are fixed to. Fixing from 
$\kf\rightarrow K\pi$, a partial width with a small experimental 
uncertainty \cite{tot}, yields a total $\df$ width of $15-16$ MeV.
Fixing from $\kf\rightarrow \ks\pi$ has the advantage that the dominant decay
$\df\rightarrow D\pi$ has almost exactly the same momentum $p$, so that
$\Gamma(\df\rightarrow D\pi)/\Gamma(\kf\rightarrow \ks\pi) = 3/4$ from 
heavy quark symmetry independent of the assumed form factor. 
Here the total $\df$ width is $36-37$ MeV. Since we have not estimated 
$\df\rightarrow \ds(\pi\pi)_S$ due to lack of experimental data from 
K--mesons it appears likely that $\df$ cannot be interpreted as $\dex$
based on its total width, although the possibility cannot be eliminated.

The decays of $\dt$ are also estimated in Table 2. 
The total estimated width of $\dt$ is $6-7$ or $17-18$ MeV
depending on whether we fix respectively from $\kf\rightarrow K\pi$ or  
$\kf\rightarrow \ks\pi$. Since we cannot estimate
$\dt\rightarrow D(\pi\pi)_S,\; \ds(\pi\pi)_S$, which have substantial phase 
space,
the balance of probability is that the total $\dt$ width is not consistent 
with
the DELPHI value.

\vspace{.5cm}

The conclusion of this section is that all interpretations of $\dex$ have too
large total widths, except possibly $\df$ and $\dt$, of which $\dt$ appears 
to be the narrowest
candidate.

\section{Interpreting $\dex$: Nodal Gaussian form factor}


Based on the $^3P_0$ model decay amplitude, we postulate the 
 nodal Gaussian form factor \cite{kokoski87,biceps}

\beqn \plabel{node}
 F^{j_q\; j_q^{'}}_{j_h\; \ell} (p^2) = F^{j_q\; j_q^{'}}_{j_h\; \ell} (0)\;
\left(1-\frac{2}{15}\frac{p^2}{\bt^2}\right)^2 \;\exp(-\frac{p^2}{6 \beta^2})
\eeqn
at the cost of introducing an extra parameter $\bt$.
The experimental motivation for this form factor 
is that the experimental ratio 
$\Gamma(\ksl\rightarrow K\pi)/\;\Gamma(\ksl\rightarrow \ks\pi) < 0.16$ 
\cite{pdg98} is
at least a  factor of eight smaller than the 
heavy quark symmetry prediction with a Gaussian form 
factor\footnote{Also, the
heavy quark symmetry prediction for 
$\Gamma(\ksl\rightarrow K\pi)/\;\Gamma(\k\rightarrow\ks\pi)$
with a Gaussian form factor is five times larger than experiment.}. 
This indicates
the need for a form factor which can additionally suppress 
$\ksl\rightarrow K\pi$.
The theoretical motivation is that the radially excited $\ksl$ should 
have a node in its wave function, which would naturally translate into 
a node in the decay amplitude. A nodal Gaussian form factor (Eq. \ref{node}) 
is accordingly found in the phenomenologically successful $^3P_0$ model;
for the decay to $D\pi,\;\ds\pi$ of all interpretations of $\dex$ discussed 
in the previous 
section, except $\df$ and $\dt$. For these interpretations we perform a 
search for decays 
to $D\pi$ and $\ds\pi$ consistent with
the DELPHI bound, using the methods of the previous section. 
Only successful searches are highlighted.

\vspace{.5cm}{\bf $\dpp$, using $\ksl$}\vspace{.1cm}

Using Eqs. \ref{hq} and \ref{node}, the 
ratio $\Gamma(\ksl\rightarrow K\pi)/\;\Gamma(\ksl\rightarrow \ks\pi)$ and 
width $\Gamma(\ksl\rightarrow K\pi)$ from experiment \cite{pdg98,tot},
we determine $F^{\frac{1}{2}\frac{1}{2}}_{1\; 1}(0)$ and 
$0.21 \leq \bt \leq 0.25$ GeV,
where the two extrema of the range has the advantage that they allow
consistency with other experimental 
data\footnote{$\Gamma(\ksl\rightarrow K\pi)/\;\Gamma(\k\rightarrow \ks\pi)
 = 0.13,\; 0.11$
for $\bt=0.21,\; 0.25$ GeV respectively, versus an experimental value of 
0.14 \protect\cite{tot}.}.
From Eqs. \ref{hq} and \ref{node} we estimate  
$\Gamma(\dpp\rightarrow \ds\pi)$, which is substantial ($\sim 100$ MeV) 
for most of the allowed $\bt$ range.
However, for the lower extremum $\Gamma(\dpp\rightarrow \ds\pi)$ is as low as
$10$ MeV. Assuming flavour SU(3) symmetry, we can also estimate
$\Gamma(\dpp\rightarrow \ds\eta)=20,\; 26$ MeV at the lower extremum 
\cite{eta}. Fixing from 
$\Gamma(\ksl\rightarrow K\pi)$ we find that the sum of the 
decays to $\ds\pi,\;\ds\eta$ and $\dss K$ can be as low as $7.9$ MeV and 
consistent with the DELPHI total width
for $0.13 \leq \bt \leq 0.20$, a region that is disjoint, but tantalizingly 
close, to the preferred region $0.21 \leq \bt \leq 0.25$ GeV. 
$\dpp$ should hence be considered too wide to be in agreement with the 
DELPHI width, 
although this depends sensitively on the experimental data on 
$\ksl\rightarrow K\pi,\; \ks\pi$.

\vspace{.5cm}{\bf $\dpp$, using $\ksh$}\vspace{.1cm}

The ratio $\Gamma(\ksh\rightarrow K\pi)/\;\Gamma(\ksh\rightarrow \ks\pi)$ 
is $1.30^{+0.23}_{-0.14}$ 
(or $2.8\pm 1.1$ directly from the LASS data) \cite{pdg98}.
This ratio, together with $\Gamma(\ksh\rightarrow K\pi)$ \cite{tot} is used
to estimate $\Gamma(\dpp\rightarrow \ds\pi)$, and within SU(3) symmetry
$\Gamma(\dpp\rightarrow \ds\eta,\;\dss K)$. We are able to find a total 
$\ds\pi, \;\ds\eta$ and
$\dss K$ width of less than 15 MeV
 only when we assume $\Gamma(\ksl\rightarrow K\pi)/\;\Gamma(\ksl\rightarrow 
\ks\pi) \geq 3.4$,
consistent, but at the very edge of the LASS error bars\footnote{At the edge 
we obtain a $\ds\pi$ width as low as 9.7 MeV. The solutions have 
$\bt= 0.15-0.16$ GeV.}. 
Consistency with the DELPHI bound is hence unnatural, but can be achieved.


\vspace{.5cm}{\bf $\dox$, using $\kz$ and $\koh$}\vspace{.1cm}

$\koh$ is interpreted as the heavy
quark symmetry partner of the $\kz$, based on the D-- to S--wave width 
ratio and the interpretation of $\kol$ as the $j_q=\frac{3}{2}$ state 
\cite{eichten93,eichten94,suzuki}.

Fixing from the ratio
$\Gamma(\kz\rightarrow K\pi)/\;\Gamma(\koh\rightarrow\ks\pi)$
and $\Gamma(\kz\rightarrow K\pi$ \cite{tot} we obtain a solution with 
$\bt = 0.19$ GeV and 
$\Gamma(\dox\rightarrow\ds\pi) = 13$ MeV. This is remarkably narrow and due to 
the amplitude ``hitting a node''. Within SU(3) we estimate
$\Gamma(\dox\rightarrow\ds\eta) = 67 - 90$ MeV \cite{eta}. The total width 
of $\dox$ is hence
inconsistent with the DELPHI value. 

In conclusion, except for $\df$ and $\dt$ which are not assumed to have a 
nodal form factor,
all interpretations of $\dex$ have too large a width, except possibly $\dpp$.
This conclusion is contingent on our inability to calculate $\dpp$ decays to
$\dz\pi,\; \dt\pi,\; D(\pi\pi)_S$ and $D\rho$ from heavy quark 
symmetry\footnote{Decays to $\dz\pi,\; D\rho$ and $\ds\rho$ can only be
estimated from current data on $\k$, the existence of which is controversial.
 Particularly,
$\Gamma(\k\rightarrow\kz\pi)=177$ MeV \protect\cite{pdg98} is a substantial 
S--wave decay 
below threshold; and should induce a huge $\dpp\rightarrow\dz\pi$ width 
since this
decay is slightly above threshold for the $\dz$ mass of ref. 
\protect\cite{ebert}.}. 

\section{Determining the $J^P$ of $\dex$ from K--meson masses}

DELPHI made the preliminary claim of an enhancement at\footnote{Obtained 
from the datum $301\pm 4\pm 10$ MeV $ =
M(\bs\pi^+\pi^-)-M(\bs)-2M(\pi)$ \protect\cite{delphib}. 
We have ignored the possibility that decay of the enhancement to 
$B\pi^+\pi^-$ is also allowed by the data \protect\cite{delphib}.}
$5905\pm 11$ MeV decaying to $\bs\pi^+\pi^-$ \cite{delphib}.
Given the similarity of this decay mode to the observation of 
$\dex\rightarrow\ds\pi^+\pi^-$ 
\cite{delphi}, we postulate that $\bex$ and $\dex$ are analogues of each other 
with different heavy quarks, and explore the consequences. 

Up to $1/m_Q$ corrections to heavy quark symmetry, we can write for the 
mass of the heavy--light meson $\bex$ and $\dex$ \cite{eichten93,eichten94}

\beqn\plabel{mass}
M({\bex}) = M(1S)_B + E + \frac{C}{m_b} \hspace{1.3cm}
M({\dex}) = M(1S)_D + E + \frac{C}{m_c}
\eeqn
where e.g. $M(1S)_B = (3M(\bs)+M(B))/4$ is the mass of the ground state.
The efficacy of using the approach in 
Eq. \ref{mass} to estimate heavy--light meson masses is seen by noting that,
 the
predictions of this approach for $D_{sJ}(2573),\; 
D_{s1}(2536),\;
B_{J}^{\ast}(5732)$ and $B_{sJ}^{\ast}(5850)$ \cite{eichten93,eichten94}
are in as good agreement with experiment as potential models 
\cite{isgur85,ebert,isgur98}.
The first set of charm and botton quark masses is taken to be $m_c=1.48$ GeV, 
$m_b=4.8$ GeV; and the second set $m_c=1.84$ GeV, 
$m_b=5.18$ GeV, following refs. \cite{eichten93,eichten94}. These two sets of 
masses include most of the range found in potential models, particularly 
those of refs. \cite{isgur85,ebert,kokoski87,hill94}. 

Using the analogous equations to Eq. \ref{mass} and following refs.
\cite{eichten93,eichten94} by fitting $\kt,$ $\kol,$ $\dts$ and 
$\doo$ according to the
newest PDG masses \cite{pdg98}, one obtains $m_s = 0.348$ GeV for set one and
$m_s = 0.433$ GeV for set two. 

There are two equations in Eq. \ref{mass}, which we solve for the two
unknowns $E$ and $C$. Substituting these values into the expression for the
mass of the K--meson analogue of $\bex$ and $\dex$, 
$\; M(1S)_K + E + \frac{C}{m_s}$, we obtain the K--meson mass
$1820\pm 60$ MeV and $1850\pm 70$ MeV for the first and second sets of
$m_s,m_c$ and $m_b$ respectively. 

In conclusion, assuming the validity of the masses of $\bex$ and $\dex$ from 
experiment, and that they are simply analogues of each other with different
heavy quarks, the lowest order correction to heavy quark symmetry 
predicts that the K--meson analogue should have a mass of  
$1800\pm 60$ MeV or $1820\pm 70$ MeV. $\kf,\ktl$ and $\kth$ are comfortably 
within,
and $\ksh$ at the edge of,  
these mass regions. Given that $\dex$ is an analogue of one of these states,
the $J^P$ of $\dex$ is $2^-$, $3^-$ or possibly $1^-$.

\section{Comments and conclusions}

It is critical to corroborate the claim by DELPHI of such a small $\dex$ 
total width. The total width is
more discriminatory than individual partial widths, e.g. with the nodal 
form factor the decay
$\dox\rightarrow \ds\pi$ is small, but the collective decay to $\ds\pi$ 
and $\ds\eta$ is substantial. 
Dominant modes are likely to be
$D\pi,\;\ds\pi,\;\ds(\pi\pi)_S$ and for some interpretations $D(\pi\pi)_S$. 
A small width for $\dex$ would put a restrictive bound on
$\ds(\pi\pi)_S$, and for some interpretations on $D(\pi\pi)_S$. This would 
be a useful input
into models.
The $J^P$ of $\dex$
can experimentally be ascertained without partial wave analysis. For 
example, of the possibilities
considered only $0^-$ and $2^-$ should have enough phase space for the 
experimentally
challenging decay to $\dz\pi$. Only $1^+,\; 0^-$ and $2^-$ decay to 
$D(\pi\pi)_S$ and do not decay to $D\pi,\; D\eta$ and $\des K$. 

If the DELPHI mass and width of $\dex$ are confirmed, it would present a 
fascinating challenge for
theory. Within heavy quark symmetry, 
the width cannot be explained by an exhaustive list of
possibilities, except possibly if the state is $\df$, $\dt$ or $\dpp$. 
However, these possibilities are inconsistent with potential model mass 
estimates. 
Moreover, if $\dex$ is either $\df$ or $\dt$  then it appears that the other
(unobserved) resonance should appear within 20 MeV of it \cite{isgur85}.
The interpretation of $\dex$ as $\dpp$ is complicated by the fact
that decay via the 
kinematically preferred route $\ds (\pi\pi)_S$ is not allowed.
Since $\dex$ is observed in $\ds\pi\pi$, this would have to arise via 
kinematically suppressed routes like $\ds\rho$ and $\dts\pi$.
Of the potentially narrow interpretations of $\dex$, $\df$ and $\dt$ are 
preferred
when the implications of the lowest order corrections to heavy quark 
symmetry on 
heavy--light meson masses are analysed.
If one insists
that potential model mass calculations are correct, $\dex$ must be $\dsp$, 
and we speculate that complicated
nodal dynamics may give rise to the experimental width. This may serve
as a sensitive probe of detailed decay dynamics, yielding tantalizing 
insight into the pair creation
mechanism, e.g. $^3P_0$ pair creation \cite{kokoski87,biceps,page97rad}.

\end{document}